\documentclass[pra,amsmath,amssymb,twocolumn]{revtex4}
\usepackage{mathrsfs}
\usepackage{amssymb}

\input epsf.tex
\usepackage{graphicx}
\usepackage{amsthm}

\begin{document}

\title{Effect of Intensity Modulator Extinction \\on Practical Quantum Key Distribution System}

\author{Jing-Zheng Huang, Zhen-Qiang Yin*, Shuang Wang, Hong-Wei Li, Wei Chen** and Zheng-Fu Han*** }

 \affiliation
 {Key Laboratory of Quantum Information,University of Science and Technology of China,Hefei, 230026, China}

 \date{\today}
\begin{abstract}
We study how the imperfection of intensity modulator effects on the security of a practical quantum key distribution system. The extinction ratio of the realistic intensity modulator is considered in our security analysis. We show that the secret key rate increases, under the practical assumption that the indeterminable noise introduced by the imperfect intensity modulator can not be controlled by the eavesdropper.
\end{abstract}
\maketitle

\section{Introduction}  \label{intro}
Quantum key distribution(QKD) is a peculiar technology that guarantees two remote parties, Alice and Bob, to share a secret key
which is prevented from eavesdropping by physical laws. Since the first QKD protocol, namely the BB4 protocol, was proposed by Bennett and Brassard in 1984\cite{BB84}, QKD has become the primary application of quantum information science. The unconditional security of QKD, which is the biggest advantage of this technology, have already been proven in both discrete-variable protocols(\cite{Theo-sec1,Theo-sec2,Theo-sec3,Theo-sec4,Theo-sec5,GLLP}) and continuous-variable protocols\cite{cv-sec6,cv-sec7,cv-sec8}. However, some assumptions in these proofs are not removable. For instance, the single-photon detectors(SPD) are usually considered to be unresponsive to bright lights in the linear mode, but it is not true. In Ref.\cite{Makarnov}, the authors experimentally prove that the SPD click can be triggered by bright light when the SPD is working in linear mode, therefore a detector-blind-attack is proposed and used to hack the commercial QKD systems successfully. For this reason, it is very important to analyze how the imperfect devices effect on the security of a practical QKD system, and the analysis of the security in practical QKD systems attracts much attention recently\cite{Prac-sec2}. For examples, the decoy state method proposed by Hwang\cite{Decoy0} has developed as a powerful tool to analyze the imperfect single-photon source and achieve the modified secret key rate\cite{Decoy1,Decoy2,Decoy3}, and the effect of imperfect phase modulator has also be studied by an equivalent model in\cite{Li-1,Li-2}. However, many imperfections are still not considered before. In this paper, we analyze the finite extinction of imperfect intensity modulator(IM) in the practical QKD system. Our simulation result shows that surprisingly, the extra noise introduced by the realistic IM reduces Eve's information and increases the secret key rate.

\section{Preliminary}
\label{sec:1}
In the BB84 protocol, Alice randomly encodes $\{0,1\}$ into the rectilinear basis $\{|H\rangle,|V\rangle\}$ or the diagonal basis $\{|+\rangle,|-\rangle\}$, where $\langle H|V\rangle=0$ and $|\pm\rangle=\frac{1}{\sqrt{2}}(|H\rangle\pm|V\rangle)$. She can produce these states by using the scheme described in Figure.1, which is applicable to the high speed QKD systems\cite{EXP-pan}.

\begin{figure}
\resizebox{1\columnwidth}{!}{%
  \includegraphics{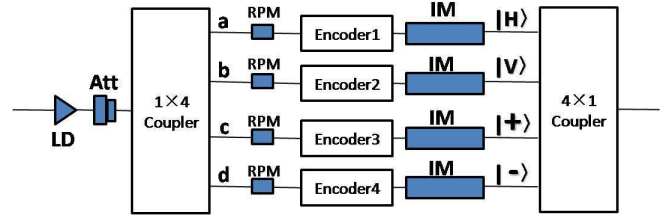}
}
\caption{The scheme diagram of producing states $|H\rangle, |V\rangle, |+\rangle and |-\rangle$. LD: Laser Diode; BS: Beam Splitter; IM: Intensity Modulator; Encoder(1-4): Polarization modulator or Mach-Zehnder interferometer, corresponding to the protocol's requirement; RPM: Random Phase Modulator. The two couplers are used for splitting and recombing the laser pulses.}
\label{fig:1}       
\end{figure}

In this scheme, the laser pulse generated by the laser diode is first attenuated to the single photon level, which can be written as a weak coherent state $|\mu\rangle$($\mu\simeq0.2$), where $\mu$ is the amplitude of light. This state is then split by a coupler into four paths, which are described as a, b, c and d. After the random phase modulating and encoding, the coherent state becomes $|\mu\rangle\rightarrow(|e^{i\theta_{a}}\frac{\mu}{2}\rangle_{a,H} + |e^{i\theta_{b}}\frac{\mu}{2}\rangle_{b,V} + |e^{i\theta_{c}}\frac{\mu}{2}\rangle_{c,+} + |e^{i\theta_{d}}\frac{\mu}{2}\rangle_{d,-})$, where
$\theta_{a,b,c,d}$ denote the additional random phases. The intensity modulators(IM) are then used to filter out the components of the state that we do not need. Depending on the electro-optic effect, we can control the attenuation of IM by supplying
an appropriate voltage on it. The state of IM is denoted as "off" when it has high attenuation and "on" when it has nearly
no attenuation. The amplitude of light is attenuated by a factor of $\sqrt{P_{off}}$ or $\sqrt{P_{on}}$ when the state of IM is off or on, and the power of the output light is $I_{off}=P_{off}I_0$ or $I_{on}=P_{on}I_0$ respectively, where $I_0 = |\frac{\mu}{2}|^2$ denotes the power of the input light.
Then the extinction ratio of IM can be defined as\cite{wiki}:
\begin{equation}                    \label{extinction}
\begin{array}{lll}
r := \frac{I_{on}}{I_{off}}.
\end{array}
\end{equation}
For simplicity, we assume the four IM have the same extinction ratio. Without loss of generality, we suppose that Alice produces the signal state $|H\rangle$. Note that only the single photon state can take part in the secret key generation\cite{GLLP}, we consider the single photon state directly. After the 4$\times$1 coupler, the single photon state at the output port can be written as (see the Appendix section for more details):
\begin{equation}                   \label{rhoh}
\begin{array}{lll}
\rho_{H} &= e^{-\frac{(P_{on}+3P_{off})\mu^2}{8}} [\frac{(P_{on}+3P_{off})\mu^2}{8}] \\
&\times (\frac{r-1}{r+3}|H\rangle\langle H| + \frac{4}{r+3}\hat{I}).
\end{array}
\end{equation}
After normalizing, the single photon part of any signal state produced by this modulation process can be generally written as:
\begin{equation}                   \label{rhos}
\begin{array}{lll}
\rho_{signal} = \frac{r-1}{r+3}\rho_{ideal} + \frac{4}{r+3}\rho_{noise}.
\end{array}
\end{equation}
Where $\rho_{ideal}$ denotes the pure signal state ($|H\rangle\langle H|$, $|V\rangle\langle V|$, $|+\rangle\langle +|$ or $|-\rangle\langle -|$), and $\rho_{noise}= \frac{1}{2}\widehat{I}$ is the density matrix of the extra noise introduced by
the finite extinction ratio of IM. Figure.2 shows the equivalent model description of the state generation process.
\begin{figure}
\resizebox{1\columnwidth}{!}{%
  \includegraphics{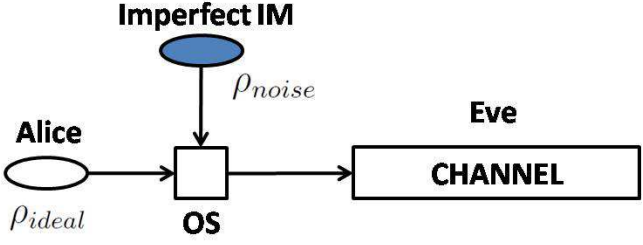}
}
\caption{The equivalent model diagram of the state generation process. OS is an optical switch that randomly routes $\rho_{ideal}$ or $\rho_{noise}$ to the channel with $\frac{r-1}{r+3}$ or $\frac{4}{r+3}$ probability. $\rho_{ideal}$, $\rho_{noise}$ and $r$ are defined in the text.}
\label{fig:1}       
\end{figure}

\section {Security Proof}
\label{sec:2}
In this section, we prove the security of the practical QKD system using the state generation scheme described in Fig.1.
We first analyze the security in the case of BB84 protocol with ideal single photon state, and then generalize our result by combining the decoy state method.

\subsection{BB84 protocol with single photon state}

The secret key rate of BB84 protocol with ideal single photon state can be written as\cite{GLLP}:
\begin{equation}                    \label{Rsingle}
\begin{array}{lll}
R_{s}= 1-H(e_1)-H(e_1).
\end{array}
\end{equation}
Where $e_1$ is the quantum error bit rate(QBER) caused by the noises in the channel and detectors,
and $H(x) = -xlog_{2}x - (1-x)log_{2}(1-x)$ is the Shannon entropy function. In the security analysis, all of the errors
are considered to be introduced by Eve when she tries to achieve secret key information from the signal state.
Equation (\ref{Rsingle}) shows that a fraction $H(e_1)$ of the sifted key is sacrificed to perform error correction, and the same amount of information is sacrificed to perform privacy amplification\cite{GLLP}.

As we have analyzed in section 2, the imperfection of IM introduce extra
noises, therefore the QBER should be modified as:
\begin{equation}                    \label{E1'}
\begin{array}{lll}
e_{1}^{'}=e_{1}\cdot\frac{r-1}{r+3}+\frac{1}{2}\cdot\frac{4}{r+3} = (1-\frac{4}{r+3})e_1+\frac{2}{r+3}.
\end{array}
\end{equation}
The second part of the right hand side of this equation is due to the 50$\%$ error rate introduced by $\rho_{noise}$. Note that
the identity matrix remains unchanged under all unitary operations, so that Eve can not achieve any useful information by performing any operation on it. Therefore this part of errors are impossible to be introduced by Eve.
For this reason, only a fraction $(1-\frac{4}{r+3})$ of the sifted key need to perform the privacy amplification process\cite{GLLP}.
We rewrite $e_1$ as a function of $e_1'$ from Eq.(\ref{E1'}):
\begin{equation}                    \label{E1}
\begin{array}{lll}
e_1=\frac{e_{1}^{'}-p}{1-2p}.
\end{array}
\end{equation}
Where $p \equiv \frac{2}{r+3}$. The modified secret key rate can now be written as follow:
\begin{equation}                    \label{Rsingle'}
\begin{array}{lll}
R_{s}^{'}=1-H(e_1^{'})-(1-2p)H(e_1)\\
~~~=1-H(e_1^{'})-(1-2p)H(\frac{e_{1}^{'}-p}{1-2p}).
\end{array}
\end{equation}

\subsection{BB84 protocol with decoy state method}
For preventing the photon number splitting(PNS) attack, decoy state method is an essential tool in practical QKD system\cite{Decoy1,Decoy2}.
The secret key rate of the BB84 protocol with decoy state method is given by\cite{Decoy2}:
\begin{equation}                    \label{Rdecoy}
\begin{array}{lll}
R_{d}=q\{-Q_{\mu}f(E_{\mu})H(E_{\mu})+Q_1[1-H(e_1^U)]\};
\end{array}
\end{equation}
\begin{equation}                    \label{Qu}
\begin{array}{lll}
Q_\mu=\sum_{i=0}^{\infty}Y_{i}e^{-\mu}\frac{\mu^{i}}{i !},\\
Q_\mu E_\mu=\sum_{i=0}^{\infty}Q_ie_i.
\end{array}
\end{equation}
Where q is the efficiency of the protocol which equals $\frac{1}{2}$ in the BB84 protocol,\\
$Y_i$ is the yield of an i-photon sate,\\
$Q_\mu$ is the gain of the signal states,\\
$E_\mu$ is the average QBER,\\
$e_i$ is the QBER caused by the i-photon state,\\
$e_1^U$ is the upper bound of $e_1$,\\
f(x) is the bidirectional error correction efficiency as a function of error bit rate\cite{bid}. $e_1^U$ can be estimated by\cite{Decoy3}:
\begin{equation}                    \label{errorup}
\begin{array}{lll}
e_1^U = \frac{e_0Y_0+e_{detect}\eta}{Y_0+\eta}.
\end{array}
\end{equation}
Where $\eta$ is the detection efficiency of the QKD system, $e_0 = \frac{1}{2}$ is the error bit rate caused by the vacuum state and $e_{detect}$ is the error bit rate caused by the system imperfection, which includes the error bit rate introduced by the imperfect IM. These two parameters can both be obtained from the field test experiments(e.g. \cite{QCC}).

Using Eq.(\ref{E1}), we get the error bit rate introduced by the ideal single photon state $\rho_{ideal}$ in Eq.(\ref{rhos}) as:
\begin{equation}                    \label{error}
\begin{array}{lll}
e_{detect} \rightarrow \frac{e_{detect}-p}{1-2p} \\
e_1^U \rightarrow \frac{\frac{Y_0}{2} + (e_{detect}-p)\eta/(1-2p)}{Y_0 + \eta} \equiv e_1^d.
\end{array}
\end{equation}
Using the similar method in section 3.1, we achieve the modified secret key rate of BB84 protocol with the decoy state method by considering the effect of IM extinction ratio as follow:
\begin{equation}                    \label{Rdecoy'}
\begin{array}{lll}
R_{d}^{'} = q\{-Q_{\mu}f(E_{\mu})H(E_{\mu})+Q_1(\frac{Y_0 + (1-2p)\eta}{Y_0 + \eta})H(e_1^d)\}.
\end{array}
\end{equation}

\section {Simulations}
\label{sec:3}
Fig.3 and Fig.4 show the numerical simulations of Eq.(\ref{Rsingle'}) and Eq.(\ref{Rdecoy'}) respectively. Here we assume the extinction ratio of IM to be 500(27dB), which is a reasonable value depending on the testing result of the IM in our field test experiment\cite{QCC}.
\begin{figure}
\resizebox{1\columnwidth}{!}{%
  \includegraphics{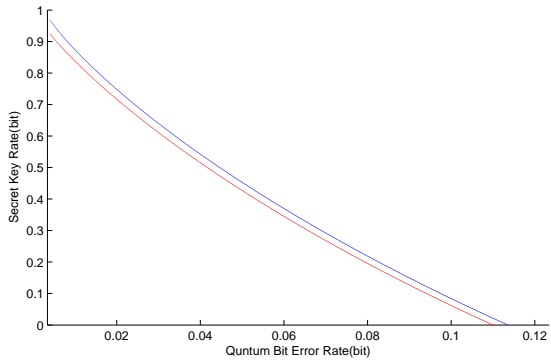}
}
\caption{The simulation result of the modified secret key rate in the case of using ideal single photon state.
The modified result in our analysis which has the maximum tolerable QBER of 11.37\% is described in blue line,
comparing to the one using the GLLP method\cite{GLLP}, which is described in red line. In this simulation,
the extinction ratio of IM is assumed to be 27dB. }
\label{fig:1}       
\end{figure}
In Fig.3, we simulate the relation between secret key rate and QBER in BB84 protocol with ideal single photon state. We
find that the maximum tolerable QBER increases to 11.37\%, comparing with 11\% when using the GLLP method.
Because the background noises caused by the imperfect IM always exist, the QBER labeled on lateral axis does not start at zero.
\begin{figure}
\resizebox{1\columnwidth}{!}{%
  \includegraphics{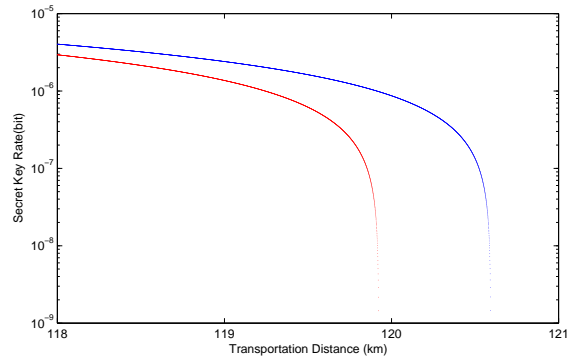}
}
\caption{The simulation result of the modified secret key rate in the case of using decoy state method.
The red line describes the optimal estimation of the secret key rate using the method in\cite{Decoy3} and experiment data in\cite{QCC}.
By considering the extra noise introduced by the imperfect IM,
the estimation of secret key rate using
the modified method in our paper which is described in blue line.
In this simulation, the extinction ratio of IM is 27dB.}
\label{fig:1}       
\end{figure}

In the case of BB84 protocol with the decoy state method, we simulate the relation between secret key rate and transportation distance
by using the key parameters for QKD experiment in\cite{QCC}. Fig.4 shows that in our analysis the maximum secure transportation distance is 120.6km, comparing to 119.9km by using the previous analysis method\cite{Decoy3}, and the secret key rates we achieved in \cite{QCC} increase $3\%$ - $4\%$(see table 1) in the modification.
\begin{table}
\caption{The secret key rates calculated by\cite{QCC} and modified by this paper, in units of kbit/s. The items in the first row denote the different communication channels of the QKD network stated in \cite{QCC}.}
\label{tab:1}       
\begin{tabular}{lllll}
\hline\noalign{\smallskip}
QKD Channels & A2R2B & A2R2C & D2R2A & E2R2A \\
\noalign{\smallskip}\hline\noalign{\smallskip}
Secret key (in \cite{QCC}) & 4.91 & 2.02 & 1.82 & 0.41 \\
Secret key (in this paper) & 5.09 & 2.09 & 1.88 & 0.43 \\
\noalign{\smallskip}\hline
\end{tabular}
\end{table}

\section {Conclusion}
\label{sec:4}
The effect of the IM extinction on practical QKD system is analyzed in paper. We surprisingly found that Eve loses more information than Alice and Bob in the process of realistic intensity modulation. We improved the lower bound of secret key rate of BB84 protocol with ideal single photon state and decoy state method respectively, and compare them to the previous security analysis in numerical simulation. Our study makes an effort to consummate the security analysis of the realistic devices in practical QKD systems. Moreover, the security of the other realistic devices can also be analyzed by the approach we derived in this paper.

\section *{Acknowledgements}
This work was supported by the National Basic Research Program of China (Grants No. 2011CBA00200 and No. 2011CB921200), National Natural Science
Foundation of China (Grants No. 60921091 and No. 61101137), and China Postdoctoral Science Foundation (Grant No. 20100480695).\\
$^*$To whom correspondence should be addressed, Email: yinzheqi@mail.ustc.edu.cn.\\
$^{**}$To whom correspondence should be addressed, Email: kooky@mail.ustc.edu.cn\\
$^{***}$To whom correspondence should be addressed, Email: zfhan@ustc.edu.cn

\section *{Appendix}
Here we derive the expressions of Eq.(\ref{rhoh}) and Eq.(\ref{rhos}) in section 2. We analyze the case of polarization encoding, which uses the $1 \times 4$ coupler described in Fig.5. The similar analysis can be also applied to the phase encoding case and obtains the same result.

\begin{figure}
\resizebox{1\columnwidth}{!}{%
  \includegraphics{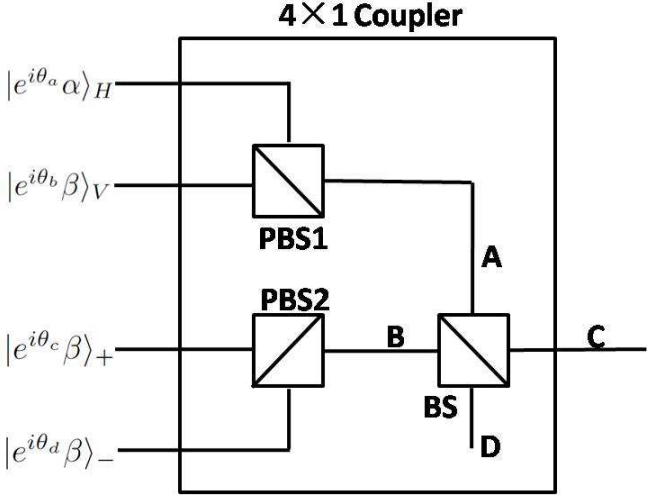}
}
\caption{The $4\times1$ coupler used in polarization encoding. PBS: Polarization Beam Splitter; BS: Beam Splitter. The four paths of the BS are denoted by A, B, C and D. The four input coherent states are defined in the text.}
\label{fig:1}       
\end{figure}
As we have mentioned in section 2, after the random phase modulating and encoding, the four coherent states input to the $4 \times 1$ coupler are $|e^{i\theta_a}\sqrt{P_{on}}\frac{\mu}{2}\rangle$, $|e^{i\theta_b}\sqrt{P_{off}}\frac{\mu}{2}\rangle$, $|e^{i\theta_c}\sqrt{P_{off}}\frac{\mu}{2}\rangle$ and $|e^{i\theta_d}\sqrt{P_{off}}\frac{\mu}{2}\rangle$, when we assume Alice produces the H state. Let us denote $\alpha = \sqrt{P_{on}}\frac{\mu}{2}$ and $\beta = \sqrt{P_{off}}\frac{\mu}{2}$. After combining by the polarization beam splitters, we get two output coherent states written in the form of creation operators acting on vacuum states:
\begin{equation}
\begin{array}{lll}
|e^{i\theta_a}\alpha\rangle_H|e^{i\theta_b}\beta\rangle_V
&\rightarrow e^{-\frac{\alpha^2+\beta^2}{2}}\sum_{n=0}^{\infty}\frac{(e^{i\theta_a}\alpha a_H^{\dag}+e^{i\theta_b}\beta a_V^{\dag})^n}{n!} |0\rangle_A \\
&= e^{-\frac{\alpha^2+\beta^2}{2}}\sum_{n=0}^{\infty}\frac{(A^{\dag})^n}{n!}|0\rangle_A, \\
|e^{i\theta_c}\beta\rangle_+|e^{i\theta_d}\beta\rangle_-
&\rightarrow e^{-\beta^2}\sum_{m=0}^{\infty}\frac{\beta^m (e^{i\theta_c}a_+^{\dag} + e^{i\theta_d}a_-^{\dag})^m}{m!}|0\rangle_B \\
&= e^{-\beta^2}\sum_{m=0}^{\infty}\frac{\beta^m (e^{i\theta_c^{'}}a_H^{\dag} + e^{i\theta_d^{'}}a_V^{\dag})^m}{m!}|0\rangle_B \\
&=e^{-\beta^2}\sum_{m=0}^{\infty}\frac{(B^{\dag})^m}{m!}|0\rangle_B.
\end{array}
\end{equation}
Where $a^{\dag}$ denotes the creation operator with the subscript of different polarizations. And we also define $e^{i\theta_c^{'}} \equiv \frac{e^{i\theta_c} + e^{i\theta_d}}{2}$ and $e^{i\theta_d^{'}} \equiv \frac{e^{i\theta_c} - e^{i\theta_d}}{2}$;
$A^{\dag} \equiv e^{i\theta_a}\alpha a_H^{\dag}+e^{i\theta_b}\beta a_V^{\dag}$ and $B^{\dag} \equiv \beta(e^{i\theta_c^{'}}a_H^{\dag} + e^{i\theta_d^{'}}a_V^{\dag})$ in the above expressions. The coherent states before inputting to the beam splitter can be written in direct product form as follows:
\begin{equation}
\begin{array}{lll}
e^{-\frac{\alpha^2+\beta^2}{2}}\sum_{n=0}^{\infty}\frac{(A^{\dag})^n}{n!}|0\rangle_A \otimes e^{-\beta^2}\sum_{m=0}^{\infty}\frac{(B^{\dag})^m}{m!}|0\rangle_B \\
= e^{-\frac{\alpha^2 + 3\beta^2}{2}} \sum_{n,m=0}^{\infty} \frac{(A^{\dag})^n(B^{\dag})^m}{n!m!}|00\rangle
\end{array}
\end{equation}
At the output ports, the beam splitter transforms each term of the above operator polynomial into:
\begin{equation}
\begin{array}{lll}
\frac{(A^{\dag})^n(B^{\dag})^m}{n!m!} &\rightarrow (\frac{1}{\sqrt{2}})^{m+n} \frac{(C_A^{\dag} + D_A^{\dag})^n (C_B^{\dag} - D_B^{\dag})^m}{n!m!}\\
&= \frac{1}{\sqrt{2}^{m+n}n!m!} \sum_{r,s=0}^{\infty}\mathbf{C}_n^r \mathbf{C}_m^s (-1)^{m-s} \\
&\times (C_A^{\dag})^r (C_B^{\dag})^s (D_A^{\dag})^{n-r} (D_B^{\dag})^{m-j}
\end{array}
\end{equation}
Where $C^{\dag}$ and $D^{\dag}$ are the operators on path C and path D in the same form of $A^{\dag}$ or $B^{\dag}$ according to their subscripts. $\mathbf{C}^r_n$ is the number of different combinations of choosing r terms at a time from n terms.

Since we only consider the single photon state, we then trace off the subspace of path D and just retain the operators that produce single photon. Thus we derive the single photon state $\rho_H$ at the output port as:
\begin{equation}
\begin{array}{lll}
\rho_H &= e^{-(\alpha^2+3\beta^2)}\sum_{m+n=1}^{\infty}(\frac{1}{\sqrt{2}^{m+n}n!m!})^2\\
&\times [(\mathbf{C}_{n}^{1}\sqrt{(n-1)!}\sqrt{m!})^2 \\
&\times (\alpha^2+3\beta^2)^{n-1}(2\beta^2)^m(\alpha^2|H\rangle\langle H| + \beta^2|V\rangle\langle V|) \\
&+ (\mathbf{C}_{m}^{1}\sqrt{(m-1)!}\sqrt{n!})^2 \\
&\times (\alpha^2+3\beta^2)^n(2\beta^2)^{m-1}\beta^2(|H\rangle\langle H| + |V\rangle\langle V|)] \\
\\
&= e^{-(\alpha^2+3\beta^2)}\sum_{m+n=1}^{\infty} \frac{1}{2^{m+n}}[\frac{n}{n!m!}(\alpha^2+3\beta^2)^{n}(2\beta^2)^m \\
&\times (\frac{\alpha^2}{\alpha^2+3\beta^2}|H\rangle\langle H| + \frac{\beta^2}{\alpha^2+3\beta^2}|V\rangle\langle V|) \\
&+ \frac{m}{m!n!}(\alpha^2+3\beta^2)^n(2\beta^2)^{m}\frac{1}{2}(|H\rangle\langle H| + V\rangle\langle V|)] \\
\\
&= e^{-(\alpha^2+3\beta^2)}\sum_{j=1}^{\infty}(\frac{1}{2})^{j}\sum_{n=0}^{j} \\
&\times \{\frac{n}{n!(j-n)!}(\alpha^2+3\beta^2)^{n}(2\beta^2)^{j-n} \\
&\times [\frac{\alpha^2-\beta^2}{\alpha^2+3\beta^2}|H\rangle\langle H| + \frac{2\beta^2}{\alpha^2+3\beta^2}
\frac{1}{2}(|H\rangle\langle H| + |V\rangle\langle V|)] \\
&+ \frac{j-n}{n!(j-n)!}(\alpha^2+3\beta^2)^{n}(2\beta^2)^{j-n}\frac{1}{2}(|H\rangle\langle H| + |V\rangle\langle V|)\} \\
\\
&\approx e^{-(\alpha^2+3\beta^2)} \sum_{j=1}^{\infty}(\frac{1}{2})^j\frac{(\alpha^2+3\beta^2)^j}{(j-1)!}(\frac{\alpha^2-\beta^2}{\alpha^2+3\beta^2}|H\rangle\langle H|)\\
&+ \sum_{j=1}^{\infty}(\frac{1}{2})^j\frac{(j-1)(\alpha^2+3\beta^2)^{j-1}2\beta^2}{(j-1)!}(\frac{\alpha^2-\beta^2}{\alpha^2+3\beta^2}|H\rangle\langle H|)\\
&+
\sum_{j=1}^{\infty}(\frac{1}{2})^j\frac{(\alpha^2+3\beta^2)^j}{(j-1)!}\frac{2\beta^2}{\alpha^2+3\beta^2}\frac{\hat{I}}{2}\\
&+
\sum_{j=1}^{\infty}(\frac{1}{2})^j\frac{(\alpha^2+3\beta^2)^{j-1}2\beta^2}{(j-1)!}
[(j-1)\frac{2\beta^2}{\alpha^2+3\beta^2}\frac{\hat{I}}{2}+\frac{\hat{I}}{2}]\\
\\
&\approx e^{-(\alpha^2+3\beta^2)} \\
&\times \sum_{j=0}^{\infty}(\frac{1}{2})^{j+1}\frac{(\alpha^2+3\beta^2)^{j+1}}{j!} \\
&\times [\frac{\alpha^2-\beta^2}{\alpha^2+3\beta^2}|H\rangle\langle H| + \frac{4\beta^2}{\alpha^2+3\beta^2}\frac{\hat{I}}{2}] \\
\\
&= e^{-\frac{\alpha^2+3\beta^2}{2}} (\frac{\alpha^2+3\beta^2}{2}) \\
&\times (\frac{\alpha^2-\beta^2}{\alpha^2+3\beta^2}|H\rangle\langle H| + \frac{4\beta^2}{\alpha^2+3\beta^2}\frac{\hat{I}}{2})\\
\\
&= N\cdot(\frac{r-1}{r+3}|H\rangle\langle H| + \frac{4}{r+3}\rho_{noise}).
\end{array}
\end{equation}
Where $\rho_{noise} \equiv \frac{1}{2}\hat{I} = \frac{|H\rangle\langle H| + |V\rangle\langle V|}{2}$, $N = e^{\frac{\alpha^2+3\beta^2}{2}}\cdot\frac{\alpha^2+3\beta^2}{2}\cdot\beta^2$ is the normalizing factor, and $r = \frac{\alpha^2}{\beta^2}$ equals to the definition of extinction ratio in Eq.(\ref{extinction}). On the fourth step of this derivation, we make an approximation by neglecting the terms have coefficient $(2\beta^2)^i$ with order higher than 1, under the assumption of $\beta \ll \alpha$. Ignoring the normalizing factor, we can derive the single photon signal state as following:
\begin{equation}
\begin{array}{lll}
\rho_{signal} = \frac{r-1}{r+3}\rho_{ideal} + \frac{4}{r+3}\rho_{noise}.
\end{array}
\end{equation}
It is exactly Eq.(\ref{rhos}) in the main text.

\end{document}